\documentclass[aps,preprint,onecolumn,eqsecnum]{revtex4}
\usepackage{dcolumn}
\usepackage{bm}
\usepackage{amssymb}
\usepackage{amsmath}
\usepackage{amsfonts}

\usepackage[]{graphicx}
\begin{document}
\bibliographystyle{prsty}
\title{ The guidance theorem of de Broglie}
\author{ Aur\'elien Drezet $^{1}$}
\address{Univ. Grenoble Alpes, CNRS, Institut N\'{e}el\\
 F-38000 Grenoble, France}
\begin{abstract}
We review some aspects of the double solution theory proposed by de Broglie at the beginning of the quantum era (i.e., in the period 1924-28).  We specifically analyze and rederive the so called guidance theorem  which is a key element of the full theory.  We compare the double solution approach to the most known pilot-wave interpretation, also known as de Broglie-Bohm or Bohmian mechanics.  We explain why de Broglie rejected the pilot wave interpretation and advocated the double solution. We also discuss some philosophical issues related to difference of strategies between de Broglie on the one side and Bohm on the other side.\\
  \begin{quote}\textit{ Le grand drame de la microphysique contemporaine a \'et\'e, vous le savez, la d\'ecouverte de la dualit\'e des ondes et des corpuscules}\cite{deBroglie}\end{quote}   
\end{abstract}
 \maketitle
\section{Introduction}
\indent Contrary to a widespread belief quantum mechanics (QM) is neither a closed nor a complete theory. 
Indeed, the standard `Copenhagen' interpretation of QM is, despite its enumerable successes, barely a catalog 
of tools and operational recipes for describing measurements and experiments made in the laboratory by physicists and engineers.
However, as it is well recognized, e.g., by J.S. Bell~\cite{Bell}, this usual interpretation says nothing about the precise definition of a quantum  measurement,  
neither does it clarify the nature of the `observed' quantum systems separated from the `observer' 
macroscopic world by the vaguely defined Heisenberg quantum/classical boundary. This lack of a clear ontological framework in QM is responsible 
for a duality which is totally foreign to the former ideal of clarity prevailing in classical physics, i.e., from Newton's period until the Einstein time.\\
\indent However, this conclusion is not forced by experimental facts and better represents a minimalist 
interpretation adapted to the experimental physicist in his lab, i.e., for all practical purposes.  However, the application of this standard interpretation to the Universe 
taken as a whole leads, as it is well known, to very strong paradoxes exemplified by the famous Schr\"{o}dinger cat and Wigner's friend contradictions.\\
\indent The aim of this present work is therefore to show that we are not obliged to accept the Copenhagen retreat but that if one really want to define a clear ontological framework
adapted to QM and then return to classical determinism in space-time, one must be prepared to modify strongly the foundation of quantum physics far beyond the aim of the
Copenhagen interpretation.\\
\indent  The present work will follow the strategy opened by the work of L. de Broglie and known as the `double solution program' (DSP) in which 
particles are represented by localized solutions (i.e., solitons) of some nonlinear-field equations evolving in the usual Minkowsky space-time~\cite{deBroglie1927,deBroglie1956}. In this DSP the particle is 
compared to a localized clock continuously phase-locked to the quantum wave guiding its path; this guiding wave 
being a solution of the  usual linear Schr\"{o}dinger, Klein-Gordon or Dirac equations. In the recent years, this theory has regained an interest in part because of the role played by the beautiful experiments initiated by Y. Couder~\footnote{We dedicate the present work to the path and memory of Y. Couder} and E. Fort concerning  walking droplets bouncing on a oil bath, and mimicking some aspects of wave-particle duality~\cite{Couder} (see also the complete review by Bush~\cite{Bush,Bush2}). Unfortunately, there are not so many available reviews concerning the DSP (see however~\cite{Fargue,Fargue2,Durt}) and very often it is only mentioned  \textit{en passant} in order to introduce the most popular `Bohmian' mechanics. In 2017, the \textit{Annales de la Fondation Louis de Broglie} published a special volume (Ann. Fond. de Broglie, \textbf{42} (2017)) acknowledging the importance of de Broglie DSP~\cite{Fargue2,Durt}. In the continuity the aim here will be to review the original DSP obtained by de Broglie in the period 1924-28. In this version~\cite{deBroglie1927} only singular waves  solutions of linear wave equations are involved.  We will discuss a beautiful theorem  obtained  by de Broglie in 1927~\cite{deBroglie1927} and called the `guidance theorem' which states how singular waves are piloted by the phase of the guiding field. Curiously this theorem is never even mentioned by Bohmians. We will also review some of the biggest issues concerning the mathematical development of the DSP and explain why the so called pilot-wave interpretation (PWI) developed by de Broglie also in 1927~\cite{deBroglie1927,Solvay} where particles are point-like objects moving in a guiding field  but without producing field singularities has been favored by Bohm in its causal interpretation of quantum mechanics~\cite{Bohm1,Bohm2}. At the difference of the DSP the PWI works not only for single particle but can also be applied to the many body problem (specially in the non-relativistic regime where the Schr\"{o}dinger equation holds). While this so called `Bohmian mechanics' can be seen as a minimalist version of the DSP de Broglie  (like Einstein who coined it `too cheap for me') never liked it and rejected this approach until the end of his life favoring, instead, the DSP. Reciprocally, Bohm~\cite{Bohm} considered the DSP as too mechanical  and too classical for explaining the major issue of quantum mechanics: i.e., quantum entanglement and nonlocality existing between  several quantum objects~\cite{Bell}. This nonlocality is predicted in the PWI and this theory is actually in complete agreement with standard quantum mechanics whereas the DSP is mainly a research program full of difficulties and presently unable to justify nonlocality. The present work will not solve this issue but it constitutes the first of a series of articles by the author devoted to the DSP and its logical development. Therefore, by reviewing some of the most important results and problems concerning the single-particle DSP the author hope to show that the DSP program could be ultimately completed and fully justified.        
\section{Prehistory of the double solution approach}
\indent We remind that de Broglie already conceived the main ideas of the DSP just after his PhD thesis of 1924~\cite{deBroglie1923,deBroglie1924}. Specifically, while in the period 1923-24 he postulated, as a Grand law of nature, the association of a local clock of pulsation $\omega_0$ (i.e., in its res frame) to any quantum particle and introduced the notion of a synchronized phase-wave accompanying its motion and its internal vibration~\cite{deBroglie1923}, it is really in 1925-1926 that he developed the concept of a singular wave-field $u(x)$ (here $x^\mu=[t,\textbf{x}]$) representing the composite wave-particle system~\cite{deBroglie1925a,deBroglie1925b,deBroglie1926} and evolving in the usual 4D space-time. Generalizing some early ideas proposed by Einstein in 1905-09 for photons~\footnote{This idea of singular photon is implicit in the special relativity and photon article of 1905 and 1917~\cite{Einstein19051,Einstein19052,Einstein1917}. Einstein,  explicitly mentioned the idea of singular photon during a conference made in Salzburg in 1909~\cite{Einstein1909}. He also introduced in the 1910-1920's the concept of ghost-field~\cite{Born1926} and planed an article about photon guiding waves for the  5$^{th}$ Solvay conference~\cite{Howard1990}} 
. This wave-field was initially supposed to be a solution of the standard d'Alembert equation 
\begin{eqnarray}\Box u(x)=[\partial_t^2-\boldsymbol{\nabla}^2]u(x)=0\end{eqnarray} valid for every positions outside a moving point singularity associated with the particle~ and incorporated into the extended wave-field. In particular, looking for a singular solution associated with uniform motion and analyzing the problem in the proper reference frame $R_0$ where the particle is at rest  he found the monopolar solution
\begin{eqnarray}
u(t_0,r_0)=e^{-i\omega_0t_0}\frac{\cos{(\omega_0r_0)}}{4\pi r_0}
\label{1}
\end{eqnarray} with $\omega_0:=m$ the Compton pulsation associated with the rest mass of the particle and $r_0$ a radius going from the singularity to an observation point in $R_0$. Of course in the Lorentzian laboratory frame $R$ where the particle is moving with a uniform velocity $\textbf{v}$ the scalar $u-$wave will be a Lorentz invariant but the Lorentz transformation actually modifies the space -time coordinates $t_0$ and $r_0$. This plays a key role in de Broglie wave mechanics as explained below.  Importantly, de Broglie selected the stationary solution  (i.e., separating space and time) corresponding to a half-half separation into retarded and advanced waves, i.e., $G_{\omega_0}(r_0)=\frac{\cos{(\omega_0r_0)}}{4\pi r_0}=\frac{1}{2}[\frac{e^{i\omega_0r_0}}{4\pi r_0}+\frac{e^{-i\omega_0r_0}}{4\pi r_0}]$ (this is a time-symmetric Green function of the Helmholtz equation $[\omega_0^2 +\boldsymbol{\nabla}_0^2]G_{\omega_0}(r_0)=-\delta^3(\mathbf{x_0})$). This is fundamental because it leads to the stability of the microobject (since the system energy radiation losses are exactly compensated by the converging  advanced waves) and at the same time implies a time-symmetric causality which is reminiscent of early ideas by Tetrode and Page~\cite{Tetrode,Page} for explaining the stability of atomic orbits (such ideas were later resurrected by Fokker,\cite{Fokker}, Feynman and Wheeler in their absorber theory~\cite{WF}, by Hoyle and Narlikar for cosmological purposes \cite{Hoyle}, and by Costa de Beauregard for explaining nonlocality and the EPR paradox with retrocausality~\cite{Costa1,Costa2}). We emphasize that all this was made before Schr\"odinger even introduced his equation. In subsequent works de Broglie \cite{deBroglie1927,deBroglie1928} considered that the u-field should better obey the so called Klein-Gordon equation~\cite{deBroglie1926b} \begin{eqnarray}\Box u(x)=-\omega_0^2 u(x)\end{eqnarray} discovered by him and many others~\cite{Kragh1984}. This equation admits the simple monopolar solution 
\begin{eqnarray}
u(t_0,r_0)=e^{-i\omega_0t_0}\frac{1}{4 \pi r_0}
\label{2}
\end{eqnarray} as well as several other ones. Indeed, the Klein-Gordon equation admits constrained monopolar solutions with pulsation $\omega$ (defined in the rest frame of the particle and associated with the local clock of the particle):
 \begin{eqnarray}
u(t_0,r_0)=e^{-i\omega t_0}\frac{\cos{(\sqrt{(\omega^2-\omega_0^2)}r_0)}}{4\pi r_0} &\textrm{for } \omega\geq\omega_0 \nonumber\\
u(t_0,r_0)=e^{-i\omega t_0}\frac{e^{-\sqrt{(\omega_0^2-\omega^2)}r_0}}{4\pi r_0} &\textrm{for  } \omega_0\geq\omega.
\end{eqnarray}
 For reasons which will be exposed elsewhere we believe that the original guess was more meaningful (for the moment it is enough to say that the d'Alembert equation doesn't depend of the proper mass $\omega_0$ and is therefore more universal). Moreover, in the present article we will consider the general case. Importantly, for both Eq.~\ref{1} and \ref{2} the phase $\varphi=-\omega_0t_0$ reads in the laboratory frame (where the particle moves at the uniform velocity $\textbf{v}$) as $\varphi(x)=-kx=-\omega t+\mathbf{k}\cdot\mathbf{x}$ with $k^\mu=[\omega=\gamma\omega_0,\mathbf{k}=\omega\textbf{v}]$ and $\gamma=1/\sqrt{(1-\textbf{v}^2)}$. From this we deduce the dispersion relation $kk=\omega^2-\mathbf{k}^2=\omega_0^2$ reminiscent of the Klein-Gordon equation  for the phase-wave $\Psi(x)=e^{i\varphi(x)}$ satisfying
\begin{eqnarray}\Box \Psi(x)=-\omega_0^2 \Psi(x)\end{eqnarray} even if $u$ itself obeys $\Box u(x)=0$. The key findings of de Broglie was to observe that if we evaluate the phase at the particle location $z=[t,\textbf{z}(t)=\textbf{v}t]$ we have \begin{eqnarray}\varphi(z)=-\omega_0t_0=-\omega_1t\end{eqnarray} with $\omega_1=\omega_0 \gamma^{-1}\neq\omega$ is the clock pulsation of the particle as seen from the laboratory frame. The internal clock of the particle is thus synchronized with the phase of the monochromatic $\Psi$-wave which is also locally the phase of the $u$-wave. Moreover, introducing the proper time $\tau$ (i.e., $d\tau=\sqrt{dxdx}=dt\gamma^{-1}$) for the particle we have 
\begin{eqnarray}
\frac{d\varphi(z(\tau))}{d\tau}=\frac{dz(\tau)}{d\tau}\partial_z \varphi(z(\tau))=-\omega_0,\nonumber\\ \textrm{with  } \frac{dz(\tau)}{d\tau}=-\frac{\partial_z \varphi(z(\tau))}{\omega_0}.
\end{eqnarray} 
This is the guidance condition that de Broglie saw as a key feature of the DSP and PWI for understanding wave-particle duality. We point out that in his first work~\cite{deBroglie1924b1,deBroglie1924b2} de Broglie used instead Rayleigh's formula $\textbf{v}=\frac{\partial \omega}{\partial \textbf{k}}$ to define the particle as a wave-packet which is by essence a polychromatic and dispersive structure. However, the DSP used in Eqs.~\ref{1},\ref{2} favors a model of monochromatic singular-fields. While a plane wave expansion of such fields is possible, $u(x)$ is rigorously not everywhere a solution of the homogeneous d'Alembert or Klein-Gordon equations because of the presence of the singularity. This induces the presence of a bound near-field which in turn modifies the dispersion relation of the plane waves $\omega_0^2\neq k^2$ appearing in the Green modal expansion~\cite{Jackson} $u(t_0,r_0)=\int \frac{d^3 \textbf{k}}{(2\pi)^3} e^{i(\textbf{k}\textbf{x}_0-\omega_0t_0)}\mathcal{P}[\frac{1}{\textbf{k}^2-\omega_0^2}]$ associated with Eq.~\ref{1} ($\mathcal{P}[...]$ denotes the principal value). Therefore, even though Rayleigh's formula can be applied in the DSP and in PWI the physical meaning is a bit different from the usual dispersion theory.\\    
\section{The theory of 1927}
\indent After this condensed summary of the early ideas about the DSP we go to the work of 1927-28 \cite{deBroglie1927,deBroglie1928} where de Broglie attempted to extend the DSP to non-monochromatic guiding waves  $\Psi(x)$ in free space and in external fields, i.e., in order to describe particle interactions with potentials and obstacles. For this purpose we introduce the more general Klein-Gordon equation for the $\Psi-$wave $\in \mathbb{C}$ in presence of external fields:     
 \begin{eqnarray}
(\partial+ieA(x))(\partial+ieA(x))\Psi(x)
=-(\chi(x)+\omega_0^2)\Psi(x)\nonumber\\ \label{3}
\end{eqnarray} 
where $e$ is the electric charge, $A^\mu(x)=[V,\textbf{A}]$ an external electromagnetic vector potential, and $\chi(x)$ an external scalar potential. This wave equation contains the Schr\"odinger equation in the non relativistic regime and it was already recognized at that time by Max Born and others that the continuous $\Psi$ wave must be interpreted statistically. Actually, this idea was also explicit in de Broglie work since 1924. However, at the difference of Born~\cite{Born1926} de Broglie conceived the $\Psi$-wave as a dynamic guiding agent for the particle, i.e., having both an ontic and epistemic status.  We here recognize the key ideas of the PWI~\cite{deBroglie1927} which de Broglie developed further for the $5^{th}$ Solvay conference~\cite{Solvay} (see also~\cite{deBroglie1930}). In the PWI the particle is a point-like object immersed in the $\Psi-$field guiding its motion and at the same time determining the probability evolution and conservation (i.e., like in classical statistical physics). Yet, in the PWI the precise meaning of the $\Psi-$wave is unclear. \\
\indent  Moreover, we introduce on top of the DSP beside the continuous $\Psi$-wave the singular $u$-wave $\in \mathbb{C}$ presenting a typically moving singularity and representing a more complete description of the corpuscle. For the sake of generality we write   
\begin{eqnarray}
(\partial+ieA(x))(\partial+ieA(x))u(x)
=-(\chi(x)+\Omega^2)u(x)\nonumber\\ \label{4}
\end{eqnarray} where $\Omega$ is not necessary identical to $\omega_0$ (in 1927 and later writings~\cite{deBroglie1927,deBroglie1956,deBroglie1957} de Broglie considered only the case $\Omega=\omega_0$).
Introducing the de Broglie-Madelung~\cite{deBroglie1925a,deBroglie1926,Madelung1927} polar representation  $\Psi(x)=a(x)e^{iS(x)}$  and  $u(x)=f(x)e^{i\varphi(x)}$ with $a,S,f,\varphi\in \mathbb{R}$ we deduce (from Eq.~\ref{3} and \ref{4}) two sets of equations. The first one reads
\begin{eqnarray}
(\partial \varphi(x)+eA(x))^2=\Omega^2+\frac{\Box f(x)}{f(x)}+\chi(x)\nonumber\\
(\partial S(x)+eA(x))^2=\omega_0^2+\frac{\Box a(x)}{a(x)}+\chi(x)\label{5}
\end{eqnarray} and these formulas are reminiscent of Hamilton-Jacobi or Euler hydrodynamical equations for fluids. The main difference~\footnote{We also emphasize that the phase $\varphi$ (the same is true for $S$) is not univocally defined near a vortex and we have around any closed loop $C$ the quantization $\oint_{(C)}dx\partial\varphi=2\pi n$ with $n$ and integer (this is reminiscent of the Bohr-Sommerfeld quantization rules).} with classical physics being that Eq.~\ref{5} contains quantum potentials $\frac{\Box a }{a}$ and  $\frac{\Box f }{f}$ curving the paths in unusual ways. In the PWI only the $\Psi-$wave is considered and the quantum `Bohmian' potential is usually associated with $\frac{\Box a }{a}:=Q_\Psi$.\\ 
\indent The second set of equations reads 
 \begin{eqnarray}
\partial[f^2(x)(\partial \varphi(x)+eA(x))]=0 \nonumber\\ \partial[a^2(x)(\partial S(x)+eA(x))]=0 \label{7}
\end{eqnarray} and the formulas are reminiscent of conservation laws for relativistic fluids with density $f^2$ and $a^2$. These relations can equivalently be written as 
\begin{eqnarray}
v_u(x)\partial\log{(f^2(x))}=\frac{d}{d\tau}\log{(f^2(x))}=\frac{\partial(\partial \varphi(x)+eA(x))}{\sqrt{[(\partial \varphi(x)+eA(x))^2]}} 
\label{8a}
\end{eqnarray}
and
\begin{eqnarray}
v_\Psi(x)\partial\log{(a^2(x))}=\frac{d}{d\tau}\log{(a^2(x))}=\frac{\partial(\partial S(x)+eA(x))}{\sqrt{[(\partial S(x)+eA(x))^2]}} \label{8b}
\end{eqnarray}
 with $v_u(x)=-\frac{\partial \varphi(x)+eA(x)}{\sqrt{[(\partial \varphi(x)+eA(x))^2]}}$ and $v_\Psi(x)=-\frac{\partial S(x)+eA(x)}{\sqrt{[(\partial S(x)+eA(x))^2]}}$ 
two unit 4-vectors associated with the local velocity of the relativistic fluids (the operators $v_u(x)\partial=\frac{d}{d\tau}$ and $v_\Psi(x)\partial=\frac{d}{d\tau}$ define Lagrangian derivatives in these fluids with proper times along the flow lines).\\
\indent The characteristic curves associated with the flow in the two fluids allow us to introduce a set of trajectories or paths given by the equations $v_u(x)=\frac{d}{d\tau} x_u(\tau)$ and $v_\Psi(x)=\frac{d}{d\tau} x_\Psi(\tau)$. At that stage these paths are not associated with a particle but are mere properties of the continuous fluids.\\
\indent In the PWI we identify $x_\Psi(\tau)$ to particle trajectories and Eq.~\ref{7} is reminiscent of the current conservation law
$\partial_\mu J_\Psi^\mu (x)=0$ where the current is given by 
\begin{eqnarray}
J_\Psi(x)=\frac{i}{2\omega_0}\Psi^\ast(x)\stackrel{\textstyle\leftrightarrow}{\rm D}\Psi(x)
=-\frac{a^2(x)}{\omega_0}(\partial S(x)+eA(x))\nonumber\\
=\frac{a^2(x)}{\omega_0}\sqrt{[\omega_0^2+Q_\Psi(x)+\chi(x)]}v_\Psi(\tau).\label{current}
\end{eqnarray}  
This relation~\footnote{where: $\psi_1(x) \stackrel{\textstyle\leftrightarrow}{\rm D}_{\mu}\psi_2(x):=\psi_1(x)D_{\mu}\psi_2(x)-\psi_2(x)D^\ast_{\mu}\psi_1(x).$}  plays a fundamental role for interpreting probabilities and electric current in scalar QED. In the PWI we can identify $\rho_0(x)=a^2(x)\sqrt{[1+\frac{(Q_\Psi(x)+\chi(x))}{\omega_0}^2]}$ with a comoving density of probability in the rest-frame of the particle.  A clear interpretation is done in the non-relativistic regime where $J_\Psi^0\simeq \rho_0(x)\simeq a^2(x):=\Psi(x)^\ast\Psi(x)$ is identical with the quantum probability density given by Born's rule for finding a particle in an elementary 3D volume around $\mathbf{x}$ at time $t$. Moreover, in the PWI this probability is associated with ignorance a la Maxwell-Boltzmann (see~\cite{Drezet1,Drezet2} for a review) and is not a fundamental or genuine property  of a somehow mysterious and non deterministic world. We emphasize that  de Broglie  initially developed the PWI in the context of the Klein-Gordon equation for a single particle. However the theory is difficult to interpret generally because $J_\Psi$ is not necessarily a time-like and future oriented  4-vector. Therefore, during the Solvay congress of 1927~\cite{Solvay} de Broglie presented a non relativistic version  of the PWI adapted  to Schr\"{o}dinger equation for the many-body problem. It is this theory which is nowadays known as Bohmian mechanics~\footnote{It should thus be clear that de Broglie is the only creator of PWI. In the same way as we can not say that Laplace invented Newtonian Mechanics the PWI should better be called  `deBroglian' mechanics.}.\\       
\indent Now, after this reminder about the PWI we go back to the DSP.  Following de Broglie~\cite{deBroglie1957} the general principle of the DSP states:\\
\begin{quote}\textit{To every regular solution $\Psi(x)=a(x)e^{iS(x)}$ of Eq.~\ref{3} corresponds a singular solution $u(x)=f(x)e^{i\varphi(x)}$ of Eq.~\ref{4} having the same phase $\varphi(x)=S(x)$, but with an amplitude $f(x)$ involving a generally moving point singularity $z(\tau)$ representing the particle.}\end{quote}
 The relation $\varphi(x)=S(x)$ was called  `phase-harmony', `phase-matching', `phase-locking' or `phase-tuning' condition by de Broglie. 
Comparing this principle with Eq.~\ref{5} implies the strong constraint   
  \begin{eqnarray}
\Omega^2+\frac{\Box f(x)}{f(x)}=\omega_0^2+\frac{\Box a(x)}{a(x)}=(\partial S(x)+eA(x))^2-\chi(x)\nonumber\\
\label{6}
\end{eqnarray} which is supposed valid for every positions outside the singularity (i.e., if $x\neq z(\tau)$  $\forall\tau$). Furthermore, by introducing the definition $F(x)=f(x)/a(x)$  we equivalently deduce 
\begin{eqnarray}
\Box F(x)+2\partial \log{a(x)}\partial F(x)-(\omega_0^2-\Omega^2)F(x)= 0 \label{6b}
\end{eqnarray} 
which shows that the $F$-field depends on the $a-$field.\\
\indent Moreover, using the phase-matching condition we get $v_u(x)=v_\Psi(x)=\frac{d}{d\tau} x(\tau)$ where $x(\tau)=x_u(\tau)=x_\Psi(\tau)$ defines common trajectories labeled by a proper time $\tau$. For the present studies we limit ourselves to the case  $(\partial S(x)+eA(x))^2\geq 0$ and consequently the fluid velocity is time-like (we have also $v_u(x)^2=1$ and $v_\Psi(x)^2=1$). This is important since the Klein-Gordon equation admits also trajectories with space-like segments, i.e., tachyonic fluid motions which are difficult to interpret in the DSP (even though a self consistent PWI  can be proposed for this tachyonic cases as well~\footnote{This will be discussed in a subsequent article.}).\\
\indent  Furthermore, since $v_u(x)=v_\Psi(x)$ and $\Box \varphi(x)=\Box S(x)$ the two Eqs.~\ref{8a} and \ref{8b} can be combined together to give: 
\begin{eqnarray}
\frac{d}{d\tau}\log{[F^2(x)]}=0.\label{9}
\end{eqnarray}  Eq.~\ref{9} means that the density $F^2(x)$ is transported and preserved along the trajectories during the $\tau$-evolution. The requirements for the DSP to fulfill both Eq.~\ref{6b} and Eq.~\ref{9} for every points $x$ is extremely demanding and probably impossible to satisfy rigorously.\\ 
\indent Physically speaking Eq.~\ref{9} seems to contradict the original motivation of the DSP. In particular, in the 1950's Francis Perrin (\cite{deBroglie1956}, chapter 18) objected to de Broglie  that such condition implies that the solitary-wave amplitude $f(x)\propto a(x)$  in general changes in time along flow-lines near the trajectory $x\sim z(\tau)$ (i.e., near the singularity). This means that the particle can generally not be considered as a stable or permanent structure in the version of the DSP presented here. \\  
\subsection{The guidance theorem and Perrin's objection}
\indent In order to be more quantitative concerning the Perrin objection we have to discuss an important  guidance theorem obtained by de Broglie already in 1927~\cite{deBroglie1927,deBroglie1953,deBroglie1956,deBroglie1957}. This theorem has a weak and strong formulation and we should discuss both of them.  Starting with Eq.~\ref{8a} we get for the singular  $u$-field
 \begin{eqnarray}
\frac{d}{dt}\log{(f^2(x))}=[\partial_t+\textbf{v}_u(x)\cdot\boldsymbol{\nabla}]\log{(f^2(x))}\nonumber\\=-\frac{\partial(\partial \varphi(x)+eA(x))}{\partial_t \varphi(x)+eV(x)} 
 \label{10}
\end{eqnarray}  with $\textbf{v}_u(x)=-\frac{\boldsymbol{\nabla}\varphi(x)-e\textbf{A}(x)}{\partial_t \varphi(x)+eV(x)}$ the 3-velocity of the $u-$fluid. Now, watching the motion in a reference frame where the singularity is practically at rest instantaneously (i.e., in a frame where the singularity motion can be analyzed non relativistically) we expect near the particle center a multipolar field amplitude \begin{eqnarray}
f(x)\simeq \frac{\alpha(x)}{R(t)^n}\label{non}\end{eqnarray} with $n$ an integer, $R(t)=|\textbf{x}-\textbf{z}(t)|$ the distance to the singularity center and $\alpha(x)$ is a smooth and regular function. This implies the relation  $[\partial_t+\frac{d\textbf{z}(t)}{dt}\cdot\boldsymbol{\nabla}]\log{(\frac{f^2(x)}{\alpha^2(x)})}=0$  and therefore in combination with Eq.~\ref{10} 
 \begin{eqnarray}
(\textbf{v}_u(x)-\frac{d\textbf{z}(t)}{dt})\cdot\boldsymbol{\nabla}\log{(\frac{f^2(x)}{\alpha^2(x)})}\nonumber\\=(\textbf{v}_u(x)-\frac{d\textbf{z}(t)}{dt})\cdot\frac{2n\hat{R}(t)}{R(t)}\nonumber\\=-\frac{\partial(\partial \varphi(x)+eA(x))}{\partial_t \varphi(x)+eV(x)}-\frac{d}{dt}\log{(\alpha^2(x))}.
 \label{11}
\end{eqnarray}In other words, since $\frac{\partial(\partial \varphi(x)+eA(x))}{\partial_t \varphi(x)+eV(x)}$ is supposed to be finite we get the condition 
\begin{eqnarray}(\textbf{v}_u(x)-\frac{d\textbf{z}(t)}{dt})\cdot\hat{R}(t)=O(R(t))\end{eqnarray} 
and thus at the limit 
\begin{eqnarray}
\textbf{v}_u(z)=\frac{d\textbf{z}(t)}{dt}\label{guidance}
\end{eqnarray}
which means that the singularity moves at the local velocity of the $u$-field at $z$. Therefore, the singularity follows one of the path $x_u(\tau)$  of the u-field flow. 
This constitutes the weak-guidance theorem: 
\begin{quote}\textit{For any singular solution $u(x)=f(x)e^{i\varphi(x)}$ of Eq.~\ref{4} associated with a moving point $z(\tau)$ and such that Eq.~\ref{non} occurs in a local reference frame associated with the singularity we have the guidance formula: $\frac{dz^\mu(\tau)}{d\tau}=\textrm{lim}_{x\rightarrow z}\{-\frac{\partial^\mu \varphi(x)+eA^\mu(x)}{\sqrt{[(\partial \varphi(x)+eA(x))^2]}}\}$.}\end{quote}
Two remarks are important here. First, we emphasize that this theorem doesn't mean that the amplitude of the field near the singularity is necessarily transported as a whole. Indeed, if we write $I_u(t,\mathbf{x})$ the right hand side of Eq.~\ref{10} we have by formal integration along a $\textbf{x}_u(t)$ line:
\begin{eqnarray}
f(t,\textbf{x}_u(t))=f(t_0,\mathbf{x}_u(t_0))e^{\frac{1}{2}\int_{t_0}^t dt'I_u(t',\mathbf{x}_u(t'))}
\end{eqnarray} where the integral is made along the $\textbf{x}_u(t)$ line between time $t_0$ and $t$. Therefore, if $I_u\neq 0$ we have in general $f(t,\textbf{x}_u(t))\neq f(t_0,\mathbf{x}_u(t_0))$ and this even for paths $x_u(\tau)$ very close to the singularity-path $z(\tau)$. Following a suggestion of G\'{e}rard Petiau in 1956-7 de Broglie used this integral formulation to derive once again the guidance theorem~\cite{deBroglie1956,deBroglie1957} (de Broglie didn't however emphasized the role of the condition $I_u\neq 0$). As a second remark, we stress that the (weak) guidance theorem is relatively robust since it only assumes  the multipolar form $f(x)\simeq \frac{\alpha(x)}{R(t)^n}$ (which  will be partially justified later) near the singularity and  doesn't even rely on the phase-harmony condition  $\varphi(x)=S(x)$ or  $\varphi(x)\simeq S(x)$, i.e., we didn't have to introduce a guiding field $\Psi$ for its derivation.\\ 
\indent Now, if we introduce the $\Psi-$field  and accept at least a first-order contact $\textbf{v}_u(x)\simeq\textbf{v}_\Psi(x)$  Eq.~\ref{guidance} reads  (and this constitutes the strong form of the guidance theorem stated by de Broglie):
\begin{eqnarray}
\textbf{v}_\Psi(z)\simeq\textbf{v}_u(z)=\frac{d\textbf{z}(t)}{dt}\label{guidancestrong}
\end{eqnarray}
i.e., we now have that the particle singularity is guided by the local velocity $\textbf{v}_\Psi(z)$ of the $\Psi$-field. 
Importantly, if in agreement with the DSP we furthermore impose a second-order contact Eq.~\ref{9} holds and we have  
\begin{eqnarray}
\frac{d}{dt}\log{(F^2(x))}=[\partial_t+\textbf{v}_\Psi(x)\cdot\boldsymbol{\nabla}]\log{(F^2(x))}=0 
 \label{9new}
\end{eqnarray} 
which by integration along a path $\mathbf{x}_\Psi(t)=\mathbf{x}_u(t)$ leads to 
\begin{eqnarray}
F(t,\textbf{x}_u(t))=F(t_0,\mathbf{x}_u(t_0))\label{notsound}
\end{eqnarray} and shows (as already stated with Eq.~\ref{9}) that the $F-$field  (with $F=f/a$) preserves its value along paths near the singularity trajectory $\textbf{z}(t)$.  We have thus the strong form of the guidance theorem:
\begin{quote}\textit{If two solutions of the wave
equations of wave mechanics are such that one of them is regular and the other one has a moving, point-like singularity and they admit the same streamlines then the singularity of
the second solution will follow one of these streamlines}\cite{deBroglie1957}\end{quote}
As already explained Eq.~\ref{notsound}  is physically not sound since, if valid, it would imply that a particle  guided by the $\Psi-$field should have and amplitude $f(x_\Psi(\tau))\propto a(x_\Psi(\tau))$ which is in general not constant along paths located near the singularity. This casts some doubts on the possibility to justify the strong guidance theorem.  Especially, as pointed out by F. Perrin to de Broglie \cite{deBroglie1956}, this is paradoxical in the case of a spreading  $\Psi$-wave (as for example with a diffracted amplitude decaying after a pinhole). Indeed, in that case  Eqs.~\ref{9},\ref{9new},\ref{notsound} would imply that the particle amplitude $F(x)$ is constant (i.e., carried by the $v_u(x)=v_\Psi(x)$ flow) near the singularity whereas the total $u-$field  amplitude $f(x)$ should decay as $a(x)$. Since $a(x)$ can take arbitrarily small value but still (in principle) induce a particle detection very far-away from the pinhole or source it is difficult to believe that the particle $u-$ field could have such a strongly decaying amplitude~\footnote{We point out that this Perrin objection motivated  the so called `tired-light' model which was introduced  to justify   the red-shift of light coming from far-away galaxy, i.e., as an alternative and exotic explanation to the cosmological expansion. In this model advocated by de Broglie~\cite{debroglietired} and others the $\propto 1/r$ spreading of the  $\Psi-$ field  induces an amplitude decay of $u-$ field near the singularity and thus of the photon energy with time.}.\\ 
\subsection{An existence proof}     
\indent In order to conclude this section about the original de Broglie DSP and the guidance theorem we go back to the justification of the weak guidance theorem and to the missing existence proof concerning the multipolar structure $f(x)\simeq \frac{\alpha(x)}{R(t)^n}$ near the singularity (this proof was not given by de Broglie but only guessed by him). For this purpose we assume that we can write $f(x)=\beta(x)G(x)$ with the hypothesis 
\begin{eqnarray}
[\partial_t+\frac{d\textbf{z}(t)}{dt}\cdot\boldsymbol{\nabla}]G(t,\mathbf{x})=0.\label{18}
\end{eqnarray} and $\beta(x)$ a regular function.
Eq.~\ref{18} can be better understood if we use the new variables $t'=t$ and $\mathbf{x}'=\mathbf{x}-\mathbf{z}(t)$  such as $G'(t',\mathbf{x}')=G(t,\mathbf{x})$. With these variables we have also $\boldsymbol{\nabla}=\boldsymbol{\nabla}'$ and $\partial_{t'}=\partial_t+\frac{d\textbf{z}(t)}{dt}\cdot\boldsymbol{\nabla}$. Therefore, Eq.~\ref{18} means $\partial_{t'}G'=0$ and thus $G'$ is independent of $t'$, i.e., $G':=g'(\mathbf{x}')=g'(\mathbf{x}-\mathbf{z}(t))$. We now go back to Eq.~\ref{5} for the $u-$wave and using the definition $f(x)=\beta(x)G(x)$ we get
\begin{eqnarray}
\Box G(x) +2\partial \log{\beta(x)}\partial G(x)=y(x)G(x) \label{19}
\end{eqnarray}   with $y(x)=(\partial \varphi(x)+eA(x))^2-\chi(x)-\frac{\Box \beta(x)}{\beta(x)}-\Omega^2$. Moreover, with the new variables $t'$ and $\mathbf{x}'$ and the properties of $G'$ we obtain 
\begin{eqnarray}
-\partial \log{\beta}\partial G=[\frac{d\textbf{z}}{dt}\partial_{t}+\boldsymbol{\nabla}]\log{\beta}\cdot\boldsymbol{\nabla}'G'. \label{19b}
\end{eqnarray} 
and
\begin{eqnarray}
-\Box G= (1-(\frac{d\textbf{z}(t)}{dt})^2)\boldsymbol{\nabla}_{||'}^2G'+\boldsymbol{\nabla}_{\bot'}^2G'-\frac{d^2\textbf{z}(t)}{dt^2}\cdot\boldsymbol{\nabla}'G'\end{eqnarray}
where $\boldsymbol{\nabla}_{||'}$ denotes the partial derivative along the $\frac{d\textbf{z}(t)}{dt}$ direction whereas $\boldsymbol{\nabla}_{\bot'}$ is associated with the direction perpendicular to the velocity. To go further, we  have to consider three approximations: First, we will neglect relativistic effects and thus write $1-(\frac{d\textbf{z}(t)}{dt})^2\simeq 1$. This actually means that we are watching the motion of the singularity in a reference frame where it is practically at rest (in the proper rest-frame we have $\frac{d\textbf{z}(t)}{dt}=0$).  Second, we write $|\boldsymbol{\nabla}'G'|\sim |G'|/l$, and  $|\boldsymbol{\nabla}'^2 G'|\sim |G'|/l^2$ with $l$ a typical length and thus we have $|\frac{d^2\textbf{z}(t)}{dt^2}\cdot\boldsymbol{\nabla}'G'/\boldsymbol{\nabla}'^2 G'|\sim |\frac{d^2\textbf{z}(t)}{dt^2}|l$.  Remarkably, following early works by M. Born and  E.~Fermi~\cite{Born1909,Fermi1922}  on the concept of rigidity in special relativity we can show that the condition for `quasi-stationarity or rigidity' of the field $G'$ reads precisely $|\frac{d^2\textbf{z}(t)}{dt^2}|l\ll 1$. Assuming this, we  deduce $-\Box G(x)\simeq\boldsymbol{\nabla}'^2G'$. As a third approximation, we neglect spatial variations of $\beta$, $\varphi$ and of the applied external fields compared to the spatial variations of $G$ (carrying the singularity) and thus write $[\frac{d\textbf{z}}{dt}\partial_{t}+\boldsymbol{\nabla}]\log{\beta}(t,\mathbf{x})\simeq [\frac{d\textbf{z}}{dt}\partial_{t}+\boldsymbol{\nabla}]\log{\beta}(t,\mathbf{z}(t)):=\mathbf{A}(t')$ and $y(t,\mathbf{x})\simeq y(t,\mathbf{z}(t)):=B(t')$. Regrouping all these approximations together we finally have the formula    
\begin{eqnarray}
\boldsymbol{\nabla}'^2G' +2\mathbf{A}(t')\cdot\boldsymbol{\nabla}'G'+B(t')G'=0  \label{19c}
\end{eqnarray} 
or equivalently
\begin{eqnarray}
(\boldsymbol{\nabla}'+\mathbf{A}(t'))^2G'+(B(t')-\mathbf{A}(t')^2)G'=0 . \label{19d}
\end{eqnarray} 
To solve Eq.~\ref{19d} we use the transformation $G'(t',\mathbf{x}')=H'(t',\mathbf{x}')e^{-\mathbf{A}(t')\cdot\mathbf{x}'}$ which leads to 
\begin{eqnarray}
\boldsymbol{\nabla}'^2H'+(B(t')-\mathbf{A}(t')^2)H'=0. \label{19e}
\end{eqnarray} 
This equation is of the Helmholtz form  and admits multipolar solutions.   For example, considering only the radial monopolar solution we get 
  \begin{eqnarray}
H'=C\frac{\cos{(\sqrt{[B-\mathbf{A}^2]\chi}r')}}{r'} \textrm{if } B-\mathbf{A}^2\geq 0\nonumber\\
H'=C\frac{e^{-\sqrt{[\mathbf{A}^2-B]}r'}}{r'} \textrm{if } B-\mathbf{A}^2\leq 0
\label{23}
\end{eqnarray}
with $C$ a constant and $r'=|\mathbf{x}'|=|\mathbf{x}-\mathbf{z}(t)|$. Moreover, all this is supposed to have a meaning only near the singularity  where $H'\simeq \frac{C}{r'}$ and $G'(t',\mathbf{x}')\simeq H'(t',\mathbf{x}')$. Therefore, we have  $G'\simeq \frac{C}{r'}$ and finally:
 \begin{eqnarray}
f(t,\mathbf{x})=\beta(t,\mathbf{x})\frac{C}{|\mathbf{x}-\mathbf{z}(t)|}
\label{24}
\end{eqnarray}
which, with the identification $\alpha=C\beta$, has the appropriate form for deriving the weak guidance theorem, i.e., Eqs.~\ref{11}-\ref{guidance}.\\
\indent Therefore, we have shown that assuming the form $f(x)=\beta(x)G(x)$ satisfying Eq.~\ref{18} the Klein-Gordon equation for the $u-$field admits singular multipolar solutions  which obey the  weak guidance theorem. Of course, this result says nothing about the function $\alpha(x)$ and whether or not it is possible to find or construct such a function. In particular, we focused our attention on the field near the singularity but the $\alpha(x)$ function could depend strongly on the boundaries located far away from the particle center, i.e., through reflections of waves generated by the singularity on the singularity it-self. This is typically what occurs for a Green function associated with a singular wave and we expect something similar here.  We believe that the previous analysis is more or less all what can deduce from the DSP without going to a more detailed description of the singularity properties or structures (i.e., obtained if we replace the singularity by a soliton or if we define precisely the localized source of the $u-$wave).  Furthermore, the present analysis of the weak guidance theorem  based only on the $u-$wave let completely open the role of the $\Psi-$wave in the DSP for guiding the particle and therefore questions the validity of the strong guidance theorem postulated  by de Broglie in his DSP\footnote{We here emphasize that Francis Fer in his doctoral thesis \cite{Fer} analyzed the guidance theorem with retarded Green functions using analogies with general relativity discussed by Vigier (i.e., in relation with works by G. Darmois, A. Einstein and A. Lichnerowicz  concerning the motion  of singularities in a metrical background~\cite{deBroglie1953,Vigier1956}).  He also attempted to demonstrate how singularities carried by a guiding wave can mathematically merge to give rise to a localized soliton associated with a nonlinear wave equation. These very interesting issues will be discussed in a subsequent work.}. \\ 
\indent Subsequently, in the 1950's, de Broglie  clearly admitted how challenging the phase-harmony condition is and suggested (without developing the idea) to relax a bit the constraints of Eq.~\ref{9} by imposing the relation $\varphi(x)\simeq S(x)$ only in the vicinity of the world-tube associated with the particle singularity $x\simeq z(\tau)$~\cite{deBroglie1956,deBroglie1953}. This is important since  the derivation of Eq.~\ref{6} only requires a first order contact between the two fluids (i.e., $\phi(x)\sim S(x)$ and $\partial\phi(x)\sim \partial S(x)$ for points near the singularity and implying $v_u(x)\simeq v_\Psi(x)$), whereas  Eq.~\ref{9} requires a second order contact (i.e., $\phi(x)\sim S(x)$, $\partial\phi(x)\sim \partial S(x)$ and, $\partial_{i,j}^2\phi(x)\sim \partial_{i,j}^2 S(x)$ for points near the singularity and leading to $\Box \varphi(x)\simeq\Box S(x)$).  \\
\indent Moreover, soon de Broglie followed a different path and after the remarks of his collaborator Jean-Pierre Vigier \cite{deBroglie1953,Vigier1956} de Broglie modified the method and basis of the DSP by including non-linearities in the wave equation for the $u$-field \cite{deBroglie1953,deBroglie1956,deBroglie1957}. The idea was to derive the existence of the particle as a localized solitonic wave-solution of a non-linear equation. More precisely, de Broglie and Vigier hoped that the presence of the non-linearity would eventually modify Eq.~\ref{9} and  lead to a locally stable singular guided $u-$wave i.e. with a local amplitude $f(z)$ not necessarily proportional to $a(z)$  (for recent reviews concerning  nonlinearity in the context of the DSP see~\cite{Fargue,Fargue2,Durt}). The strategy was very similar to the one followed by Einstein in his quest for a theory unifying gravitation and quantum mechanics (for a review see \cite{Vigier1956}). For Einstein  the geometrical field $g^{\mu\nu}(x)$ characterizing gravity should be able to generate localized objects acting as moving particles (i.e. due to the nonlinearity of general relativity). This was for example the case with the `Einstein-Rosen' bridge~\cite{Einstein} (better known as a space-time wormhole) introduced originally as a model of particles. The great vision of Einstein~\cite{Einstein} and Vigier~\cite{deBroglie1953,Vigier1956} was thus to derive quantization from a future `geometrico-dynamics' yet to be constructed. Therefore, DSP was envisioned as a part of a larger program or quest.

\section{Conclusion}
In order to conclude this review we would like to to go back to the origin of the DSP. De Broglie was strongly motivated by the success of Einstein in general relativity  and Lorentz, Abraham, Poincar\'e or Mie in electrodynamics for developing a self-consistent model of particle in the context of classical field  theory. However,  in order to account for wave-particle duality he had to introduce the notion of phase-harmony and a guidance theorem  to define singular $u$-waves guided by regular $\Psi-$waves. The PWI, which is a by product consequence of his research program, doesn't involve singular waves and accept the notion of particle as an external pattern surfing on the $\Psi-$wave. Contrarily to some claims, the fundamental reasons explaining why de Broglie renounced to his theory in 1928  are not completely related to some technical objections made by Pauli and others at the Solvay Congress (even if this context certainly played a role) but are more connected to the fact that he could not complete the DSP and  also because the PWI was for him problematic.  It is useful here to review some of the objections made by de Broglie to his own PWI (see \cite{deBroglie1953,deBroglie1930}). First of all, de Broglie had big issues with the concept of wave collapse which was discussed during the Solvay congress. Specifically, Einstein \cite{Solvay} introduced an example with a particle diffracted by a screen and pointed out that the particle detection at one location preclude the subsequent detection of any effect of the $\Psi-$wave at any other position of the screen. In other words, the wave has disappeared or `collapsed' in agreement with Heisenberg interpretation (Heisenberg and also von Neumann actually formalized  this idea in the following years). How, could we account for that in the PWI if the $\Psi-$wave is a physical agent? A collapse is indeed acting instantaneously and therefore this would  involve faster than light action at a distance and non-locality. Furthermore, with Schr\"{o}dinger the $\Psi-$wave is generally propagating in the configuration space for the $N$-particles.  It is thus generally very difficult to find a physical content to the $\Psi-$wave in the 3D space. Subsequent works by Einstein Podolsky and Rosen in 1935 and much later by Bohm and Bell stressed even more the role of nonlocality in the PWI \cite{Bohm,Bell}. However, even without going to  Bell nonlocality the problem is already present at the single particle level as we saw with Einstein example~\footnote{from an empirical point of view however it has not yet been possible to find any manifestation of single-particle nonlocality in the past despite many claims by L. Hardy and others. Actually as pointed out years ago by A. Zeilinger~\cite{Zeilinger} an hypothetical single-particle nonlocality is always a many-particles nonlocality in disguise.}. The concept of collapse is not however necessary in the PWI as it was demonstrated by Bohm: this is an old relic of the 1930's before people understood entanglement and quantum measurements. In Einstein's experiment entanglement with detectors would have the same effect as an effective collapse. Still, despite some clarifications the PWI looks very peculiar and mysterious. Some Bohmians resigned to find a better explanation and accepted a `nomological' approach which, in the end, is not really better than the Copenhagen interpretation~\footnote{De Broglie could not afford such a perspective on the PWI. For him, there is something like a deny in the PWI philosophy since we accept that an actual trajectory is modified by all the other possible paths which could have been realized but which are not.  This interaction is carried by the quantum potential $Q_\Psi$. Therefore, we can not use the $\Psi-$wave simply like a statistical and epistemic tool  but we must add an ontological content to it. The exact nature of this ontological content is however not clear for Bohmians  and thus the retreat to a nomological  approach is at best only a temporary expedient.}. However, de Broglie could not resign. He wanted a clear description of the sequence of events in the 4D space-time not in the configuration space. That was the reason why he could not advocate the PWI even though he invented it and that's why he preferred the DSP despite its uncompletion. \\
\indent  As we saw the Guidance theorem offers an interesting promise for a future DSP.  We have proven a weak form of this guidance theorem  therefore completing the historical  proof of de Broglie. However, de Broglie hoped to justify the strong form of the guidance theorem in which  the phase of the $\Psi-$wave determines the complete motion of the singularity. However, F. Perrin objection is very important in this context since (i.e., with Eq.~\ref{9}) it means that the singularity guided by a $\Psi-$ wave  would not be a permanent object.  Of course, if we don't accept the second order phase-matching condition Eq.~\ref{9} doesn't hold anymore but still  Perrin's objection is very vivid.  Indeed, if the particle amplitude in the region of the singularity is not changing proportionally to the amplitude $a(x(\tau))$ of the $\Psi-$wave it could be that the $u-$wave sometimes looses its $\Psi-$wave. This could occur in the same situation as before where a spreading $\Psi-$wave decays continuously  whereas the $u-$wave is now keeping a more or less constant value (i.e., like a soliton or a singularity). That's a very strange consequence of the DSP which must be taken seriously. This issue is very much connected to the problem of energy conservation in the DSP (and PWI) and to the previous issue concerning wave collapse. Indeed, if a particle is going through a beam-splitter the $\Psi-$wave will be separated in two branches.  If the particle in agreement with the DSP follows only one path some energy should necessarily go in the two branches if we want in a subsequent step to realize an interference experiment by reuniting the two $\Psi$-beams.  The question is thus how small should be this amount of energy in order not to make the particle unstable or to have noticeable effect which should have been already observed .  At the same time, the energy should be big enough to disturb the  subsequent motion of the particle in the interference experiment. In the PWI the magical ingredient is the quantum potential $Q_\Psi(x)$ which in general is time dependent (even if the external potential are not) and thus the particle energy $E=-\partial_t S$ is in general not constant. But what is the physical meaning of that quantum potential? What is the source of the energy giving birth to $Q_\Psi(x)$? Is this ultimately connected to quantum vacuum~\footnote{It is remarkable that the quantum potential for an harmonic oscillator in its ground state is precisely $Q_\Psi=\hbar \omega_0/2$ where $\omega_0$ is the pulsation of the oscillator and $\hbar \omega_0/2$ is also the total energy. Applied to QED we have identity between the zero-point field energy and the quantum potential~\cite{Bohm}. } and fluctuations in the zero-point field which are already responsible of Casimir effects and drive spontaneous emission of light?  It is interesting to observe that Bohm wanted to interpret this feature as the signature of a new form of information he called `active information' \cite{Bohm} and which is used by quantum systems to guide their motions. As Bohm and Hiley wrote `the basic idea of active information is that a form having very little energy enters into and directs a much greater energy'~\cite{Bohm}. Bohm used the analogy  with a radar or a radio wave signal which carries a very small amount of energy but can be used by the human receiver to direct its future motion. For Bohm  the particle has a rich inner structure able to exploits the nonlocal information about its environment to direct its path. However, de Broglie and his collaborators  like Vigier hoped in the 1950-60's to find a mechanical explanation for the existence of this $Q_\Psi(x)$ without abandoning the possibility of a clear causal description in the 4D space-time background and without hiding everything behind the label `it is nonlocal'. Bohm criticized DSP by claiming that nonlinearity of the $u-$wave could not explain the strong effect induced by the quantum potential on the particle motion and that nonlocality is an essential element in the explanation.    Remarkably, the quantum potential  $Q_\Psi=\frac{\Box a }{a}$ only depends on the form of the wave function and not on its absolute value $a=|\Psi|$. Therefore, the key ingredient is the phase $S$ which is related to $Q_\Psi$ by Eq.~\ref{5}.  However, the phase is also the key element for interpreting de Broglie internal clock and the guidance theorem, and thus it seems to me that both the point of view of de Broglie and the one of Bohm are somehow telling the same thing.  As we see the difficulties are important and all alternatives are very much demanding. De Broglie and Vigier hoped to solve these issues by introducing nonlinearities and solitons in the DSP. In future papers of this series we will show that  with nonlinearity it is indeed possible to push the DSP to its logical development.           
\section{Acknowledgments} 
We thank C\'edric  Poulain for several important comments and discussions. We thank Daniel Fargue for reminiscences concerning the pioneer role of Francis Fer in deriving the guidance theorem.   

\end{document}